\documentclass[aps,prb,reprint,groupedaddress,amsmath,amssymb,preprintnumbers,showpacs,longbibliography]{revtex4-2}
\usepackage{graphicx}
\usepackage{hyperref}
\usepackage{amsmath,bm}
\usepackage{dcolumn}
\usepackage{braket}
\usepackage{physics}
\usepackage{float}
\usepackage{xcolor}

\newcommand{\be}{\begin{equation}}
	\newcommand{\ee}{\end{equation}}
\newcommand{\ba}{\begin{eqnarray}}
	\newcommand{\ea}{\end{eqnarray}}
\newcommand{\bs}{\begin{subequations}}
	\newcommand{\es}{\end{subequations}}
\newcommand{\bw}{\begin{widetext}}
	\newcommand{\ew}{\end{widetext}}

\usepackage[final]{showlabels}
\usepackage{cleveref}
\crefname{equation}{Eq.}{Eqs.}
\crefname{figure}{Fig.}{Figs.}
\Crefname{Figure}{Fig.}{Figs.}
\crefname{table}{Table}{Tables}
\crefname{section}{Sec.}{Secs.}
\Crefname{section}{Sections}{Sections}
\crefname{subsection}{Sec.}{Sec.}
\Crefname{subsection}{Subsection}{Subsections}
\crefname{subsubsection}{subsection}{subsections}
\Crefname{subsubsection}{Subsection}{Subsections}
\crefname{paragraph}{subsection}{subsections}
\Crefname{paragraph}{Subsection}{Subsections}

\mathcode`\*="8000
{\catcode`\*\active\gdef*{\cdot}}

\begin{document}
	\title{Doping effects in high-harmonic generation from correlated systems}
	\date{\today}
	\author{Thomas Hansen and Lars Bojer Madsen}
	\affiliation{Department of Physics and Astronomy, Aarhus
		University, DK-8000 Aarhus C, Denmark}
	\begin{abstract}
		Using the one-dimensional Hubbard model, which is commonly used for describing, e.g., high-$T_c$ superconducting cuprates, we study high-harmonic generation (HHG) from doped, correlated materials. Doping is modeled by changing the number of electrons in the lattice from the conventional half-filling case. For relatively small Hubbard $U$, i.e., small electron-electron correlation, we find little to no effect of doping on the dynamics and the HHG spectra. For increasing $U$ the degree of doping has a marked effect on the dynamics and spectra. We explain these findings through the quasiparticle-based doublon-holon picture. The dynamics are separated into two types, firstly doublon and holon movement, and, secondly, doublon-holon pair creation and annihilation. Doping results in all configurations containing doublons or holons. Those quasiparticles can move at no extra cost in energy regardless of the correlation level. This motion at no energy cost increases the high-harmonic gain for low and medium harmonic orders. We discuss that in  the high-$U$ limit, antiferromagnetic ordering becomes increasingly unlikely with increasing doping rates and explain an associated drop in the high-order harmonics relative to the case of half filling.
	\end{abstract}
	\maketitle
	
	\section{Introduction}
	High-harmonic generation (HHG) is a highly non-linear, ultrafast, atto- to femtosecond ($10^{-18}$s - $10^{-15}$s), process through which ultrashort laser pulses, with high photon energy can be produced \cite{,Schubert2014,Lewenstein1994,Lein2003,Li2008,Ghimire2011}. The HHG process also allows for retrieval of spectrographic information of the generating materials at a sub-femtosecond timescale \cite{Lein2002,Torres2007,Li2008,Kraus2015,Luu2018,Silva2018}. A prominent description of the HHG process in solids originates from the Bloch-band picture and splits the dynamics into two types: intra- and interband HHG \cite{Golde2008,Vampa2014}. The intraband HHG arises from a charge carrier propagating through a curved Bloch band, the underlying acceleration results in radiation. The interband HHG is best described using the so-called 3-step model: (i) an electron is excited across a band gap, (ii) the excited electron and created hole propagate through their respective bands, resulting in intraband radiation, and (iii) the hole and electron recombine thus emitting a photon with energy given by the energy difference between the conduction and valence bands in question at the crystal momentum at the time of recombination \cite{Vampa2014}.
	
	Since the first observation of HHG from solids \cite{Ghimire2011}, it has garnered much interest, as the higher particle density as compared to gases could enable the production of higher intensity ultrafast pulses \cite{Schubert2014,Garg2016,Luu2015,You2017,Kaneshima2018,Liu2017,Luu2018,JensenPRA2022,Yamada2021,PhysRevA.96.053418,PhysRevLett.118.087403,Floss2018}. Recently, research into HHG from highly correlated materials, such as Mott insulators, has received much interest \cite{Murakami2018,Murakami2021,Silva2018}. The Hubbard model has been used in this context as it captures certain aspects of the physics in real materials \cite{Lee06,Imada98}. One such class of materials is cuprates. Cuprates contain high-temperature superconductors and therefore much research towards understanding the mechanism leading to superconductivity in cuprates and pushing their critical temperature higher has been done \cite{MMIT97,TMO04}. As a result, it is highly interesting to elucidate the transport properties of such materials, e.g., through HHG. In all cuprate materials, the on-site electron repulsion plays a significant role in the dynamics of the system. They can therefore be described as, potentially, doped Mott insulators \cite{Lee06,Imada98}. Note that the metal-insulator transition generally happens for solids with only full, empty, or half-filled bands, i.e., Mott insulators are only achieved when every band is either full, empty, or half-filled. Another group of materials, some of which can be simulated by similar Hubbard-model-based techniques, are the nearly one-dimensional organic charge-transfer salts. These salts can range from Mott insulators, through organic metals, to superconductors \cite{Bloch83}.
	
	This leads us back to the predominant model for studying correlated materials, the Hubbard model. On-site electron-electron repulsion is included through the so-called Hubbard $U$-term. This model is frequently used to study HHG from Mott insulators \cite{Silva2018,Murakami2018,Murakami2021,Udono22}. Such studies have led to a 3-step model formulated in terms of doublon and holon quasiparticles. Doublons are doubly occupied lattice sites, and holons are empty sites.  The 3 steps being: (i) A doublon-holon pair is created, (ii) the created doublon and holon propagate throughout the lattice, and (iii) the doublon-holon pair recombines under emission of a high-energy photon \cite{Murakami2021}.
	
	Studying HHG from Mott insulators requires the half-filling assumption. However, cuprates  include materials for which the highest occupied band is far from half-filled \cite{Tokura98}.  It is noted that one-dimensional cuprate chains have been synthesized with many different degrees of hole doping \cite{Chen2021}. As the degree of band filling is of extreme importance to the dynamics \cite{Bloch83} it is of interest to go beyond the assumption of half filling and address its influence on the ultrafast dynamics in correlated materials through the nonlinear process of HHG. Especially as half filling is a special case in correlated materials since the groundstate contains neither doublons nor holons for $U\rightarrow \infty$.  So here we present what is, to the best of our knowledge, the first study of HHG using the Hubbard model for non-half-filling cases. Our goal is to elucidate the following questions: (i) What effects does doping away from half filling have on the HHG spectra? (ii) How do the dynamics of non-half-filled bands differ from half-filled bands? (iii) How do the changes in the dynamics and HHG spectra relate to one another?
	 
	The paper is organized as follows. In \cref{sec:theory}, the theoretical model and numerical methods are introduced. In \cref{sec:results}, the results are presented and analyzed and, finally, in \cref{sec:conclusion} we conclude. Atomic units are used throughout unless stated otherwise.

	\section{Theoretical model and methods}\label{sec:theory}
	We work with the Hubbard model on a one-dimensional chain of $L=12$ atoms with periodic boundary conditions. We use different numbers of electrons, keeping the number of spin-up and spin-down electrons equal at all times to consider spin-neutral situations for a range of different degrees of band filling.
	
	In the presence of a driving laser pulse, the Hubbard model Hamiltonian can be expressed as \cite{Hubbard}:
	\begin{align}
		\hat{H}&=-t_0\sum_{i,\sigma}\left(e^{iaA(t)}\hat{c}_{i+1,\sigma}^\dagger \hat{c}_{i,\sigma}+\mathrm{h.c.}\right)+U\sum_{i}\hat{n}_{i,\uparrow}\hat{n}_{i,\downarrow}, \label{eq:Hamiltonian}
	\end{align}
	where $t_0$ is the nearest-neighbor hopping matrix element which, by lattice symmetry, is independent of lattice site and hopping direction. An electromagnetic field is included via Peierl's phase $e^{\pm iaA(t)}$ where $a$ is the lattice spacing and $A(t)$ is the electromagnetic vector potential treated in the electric dipole approximation, i.e., neglecting the spatial dependence in $A(t)$.  The fermionic creation and annihilation operators for an electron on site $i$ with spin $\sigma\in\left\{\uparrow,\downarrow\right\}$ are denoted by $\hat{c}_{i,\sigma}^\dagger$ and   $\hat{c}_{i,\sigma}$, respectively. The electron-electron correlation is included via the Hubbard $U$-term. $U$ will be treated as a parameter. Finally, $\hat{n}_{i,\sigma}=\hat{c}_{i,\sigma}^\dagger \hat{c}_{i,\sigma}$ is the number operator for electrons on site $i$ with spin $\sigma$. It is seen from the last term in \cref{eq:Hamiltonian} that each doublon increases the energy in the system by $U$.
	The values for lattice spacing and hopping term were picked to fit those of $\text{Sr}_\text{2}\text{CuO}_\text{4}$ \cite{Tomita2001}, as was done previously in Ref. \cite{Silva2018}. The specific values are $a=7.5589$ a.u. and $t_0=0.0191$a.u.
	
	We use a linearly polarized, $N_c=10$ cycle pulse with polarization direction along the lattice dimension. We use a $\sin^2$ envelope. The explicit form of the vector potential is
	\begin{align}
		A(t)&=A_0\cos(\omega_Lt-N_c\pi)\sin^2\left(\frac{\omega_Lt}{2N_c}\right). \label{eq:A}
	\end{align}
	The vector potential amplitude is given by $A_0={F_0}/{\omega_L}=0.194$ a.u. and the angular-frequency by $\omega_L=0.005$ a.u.$=33$ THz. The field strength, $F_0$, corresponds to a peak intensity of $3.3\cross 10^{10}$ W/cm$^2$.
	
	The model is a one-band model so, as in previous studies \cite{Silva2018,Murakami2018,Murakami2021}, we only consider intraband dynamics. Note that intraband models have been successful in describing HHG from a variety of materials \cite{You2017,Ghimire2011,Kaneshima2018,Klemke2020,Liu2017,Luu2015,Luu2018}. 
	
	The simulations are done by solving the time-dependent Schrödinger equation via the Arnoldi-Lancoz algorithm. This goes for both the imaginary time propagation used to find the initial groundstate and the real-time propagation. The imaginary time propagation continues until the energy is converged and the electron density is translationally symmetric. The latter is required by translation symmetry of the lattice. We find that the symmetry requirement is the stricter of the two when starting the imaginary time propagation from a random initial state.

\begin{figure}
		\centering
		\includegraphics[width=\linewidth]{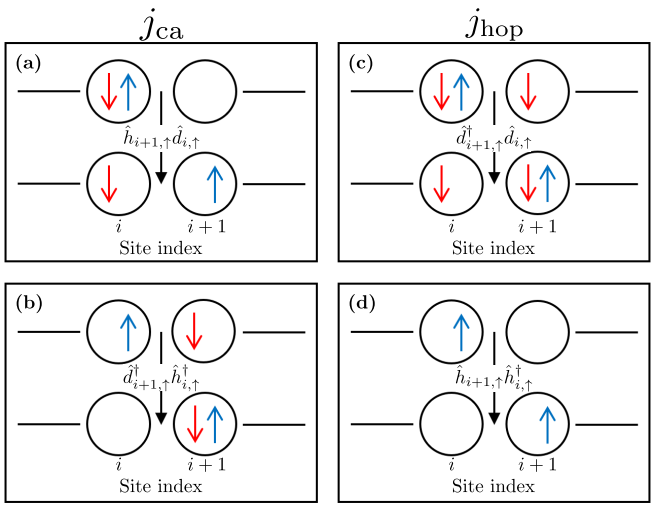}
		\caption{Illustration of the separation of the dynamics into currents, $j_\text{ca}(t)$, associated with the (a) annihilation and (b) creation of doublon-holon pairs, and into currents, $j_\text{hop}(t)$, associated with the hopping of a (c) doublon or a (d) holon. Sites are illustrated by black circles. In each panel the upper sites are the initial configuration, and the bottom the final configuration. The transition is created by the operator written in the middle of each panel. A red arrow indicate that a spin-down electron occupy the given site, similarly a blue arrow indicates that a spin-up electron occupies the given site. }
		\label{fig:illustration}
	\end{figure}

	We consider a target that is thin in the laser propagation direction. This allows us to neglect propagation and phase matching effects \cite{Gaarde2008}. In this case, the generated field is proportional to the electron acceleration, i.e., the time derivative of the current. Therefore the spectrum $S(\omega)$ is expressed as
	\begin{align}
		S(\omega)=\left|\mathcal{F}\left(\frac{d}{dt}j(t)\right)\right|^2=\left|\omega j(\omega)\right|^2, \label{eq:Spetrum}
	\end{align}
	where $j(\omega)$ is the Fourier transform ($\mathcal{F}$) of the current $j(t)=\langle \hat{j}(t)\rangle$. Here the current operator reads as \cite{Mahan}
	\begin{align}
		\hat{j}(t)=-iat_0\sum_{i,\sigma}\left(e^{iaA(t)}\hat{c}_{i+1,\sigma}^\dagger \hat{c}_{i,\sigma}-\mathrm{h.c.}\right). \label{eq:j}
	\end{align}
	In the analysis of our results, we will utilize a separation of the current into two parts. These are the currents associated with the movement of doublons and holons and the current associated with the creation or annihilation of doublon-holon pairs. This separation is in line with the earlier mentioned parallel to the 3-step model. The creation and annihilation dynamics are parallel to interband dynamics, and the doublon and holon hopping dynamics are parallel to intraband dynamics \cite{Murakami2018}. As the $U$-term of \cref{eq:Hamiltonian} effectively counts the number of doublons, this distinction is physically relevant. Following Ref. \cite{Murakami2021}, we denote these currents as $j_\text{ca}(t)=\langle \hat{j}_\text{ca}(t)\rangle$ and $j_\text{hop}(t)=\langle\hat{j}_\text{hop}(t)\rangle$ respectively. Here 'ca' refers to the creation and annihilation of doublon-holon pairs, and 'hop' refers to the current associated with the hopping of doublons and holons. The current operator can be expressed as $\hat{j}(t)=\hat{j}_\text{ca}(t)+\hat{j}_\text{hop}(t)$ with \cite{Murakami2021} 
	\begin{align}
		\hat{j}_\text{ca}(t)&=iat_0\sum_{i,\sigma}\left[e^{iaA(t)}\left(\hat{d}_{i,\sigma}^\dagger \hat{h}_{i+1,\sigma}^\dagger+\hat{h}_{i,\sigma}\hat{d}_{i+1,\sigma}\right)-\mathrm{h.c.}\right], \label{eq:j_ca}\\
		\hat{j}_\text{hop}(t)&=iat_0\sum_{i,\sigma}\left[e^{iaA(t)}\left(\hat{h}_{i,\sigma} \hat{h}_{i+1,\sigma}^\dagger+\hat{d}_{i,\sigma}^\dagger\hat{d}_{i+1,\sigma}\right)-\mathrm{h.c.}\right], \label{eq:j_hop}
	\end{align}
	where $\hat{d}_{i,\sigma}^\dagger=\hat{c}_{i,\sigma}^\dagger\hat{n}_{i,\gamma}$ and $\hat{h}_{i,\sigma}^\dagger=\hat{c}_{i,\sigma}\left(\hat{I}-\hat{n}_{i,\gamma}\right)$ are operators which create a doublon or holon, respectively, on site $i$ by creating or annihilating an electron with spin $\sigma$. Note that $\gamma$ denotes the spin, opposite to $\sigma$, and $\hat{I}$ denotes the identity operator. The operators achieve their function, in terms of hopping, creation, and annihilation, by effectively doing two things. Firstly, they check whether an electron with spin $\gamma$ is or is not on the site in question, for doublon or holon manipulation, respectively. Secondly, they create or annihilate an electron with spin $\sigma$ on the given site.  Figure \ref{fig:illustration} illustrates examples of the action of terms in Eqs.~\eqref{eq:j_ca} and \eqref{eq:j_hop} describing the hopping of a spin-up electron from site $i$ to site $i+1$.  We see from the figure how these operators (a) annihilate a doublon, (b) create a doublon, moves (c) a doublon or (d) a holon.

	In our analysis of the results, we also consider spectra generated from only $j_\text{hop}(t)$ and $j_\text{ca}(t)$, respectively. Those spectra are given by
	\begin{align}
		S_\text{hop}(\omega)=\left|\omega j_\text{hop}(\omega)\right|^2,\quad S_\text{ca}(\omega)=\left|\omega j_\text{ca}(\omega)\right|^2. \label{eq:S_ca and S_hop}
	\end{align}
	Although $S_\text{ca}(\omega)$ and $S_\text{hop}(\omega)$ can not be measured individually since $S(\omega)=\left|\omega j(\omega)\right|^2=\left|\omega j_\text{ca}(\omega)+\omega j_\text{hop}(\omega)\right|^2$, these latter spectra can still indicate the relative importance of each mechanism.

	To interpret results regarding this separation of the underlying dynamics, we utilize the measure
	\begin{align}
		D(t)=\sum_{i}\frac{\langle \hat{n}_{i,\uparrow}\hat{n}_{i,\downarrow} \rangle }{L}. \label{eq:D}
	\end{align}
	The $D$-measure is the expectation value of the $U$-term, of \cref{eq:Hamiltonian}, scaled by ${1}/{UL}$. The $D$-measure, therefore, provides a way to compare the effects of the $U$-term across different simulations. In the limit of $U\gg t_0$, the eigenstates of \cref{eq:Hamiltonian} become energetically separated into groups defined by the number of doublons in the given state. Those groups are commonly called Hubbard bands \cite{TMO04,Hubbard,MMIT97}. For high $U$, changes in $D(t)$ therefore also indicates excitation.
	
	We will also benefit from a measure that correlates more directly to the probability of doublon-holon pair creation. With that in mind, we utilize the following measure
	\begin{align}
		P_\text{af}=\Bigg\langle\sum_{\langle i,j\rangle}\frac{\hat{n}_{\uparrow,i}(\hat{I}-\hat{n}_{\downarrow,i})*(\hat{I}-\hat{n}_{\uparrow,j})\hat{n}_{\downarrow,j}}{L}\Bigg\rangle, \label{eq:Paf}
	\end{align}
	which gives the probability of observing antiferromagnetic ('af') ordering when observing two neighboring sites at random. Regarding the notation, $i$ and $j$ are nearest-neighbor lattice site indexes, denoted by $\langle i,j\rangle$. Note that antiferromagnetic ordering across two sites, depicted as the final configuration in \cref{fig:illustration} (a) and initial configuration in \cref{fig:illustration} (b),  means there is precisely one electron on each site, and that they have opposite spin. As doublon-holon pairs can only be created through transitions across antiferromagnetically ordered, nearest-neighbor sites, as depicted in \cref{fig:illustration} (b), the probability of observing sites with antiferromagnetic order is heavily correlated to the probability of doublon-holon-pair-creating transitions. 
	
	\section{Results and discussion}\label{sec:results}
	In this section, we present our results and analyzes. We consider four $U$-values: $U=0.1t_0,2t_0,5t_0,\text{ and }10t_0$, which are picked as representative values based on a more thorough parameter scan. $U$-values up to $10t_0$ have been  seen in cuprates \cite{Hybertsen89}.
	\subsection{HHG spectra}\label{sec:spectral results}
\begin{figure}
		\centering
		\includegraphics[width=\linewidth]{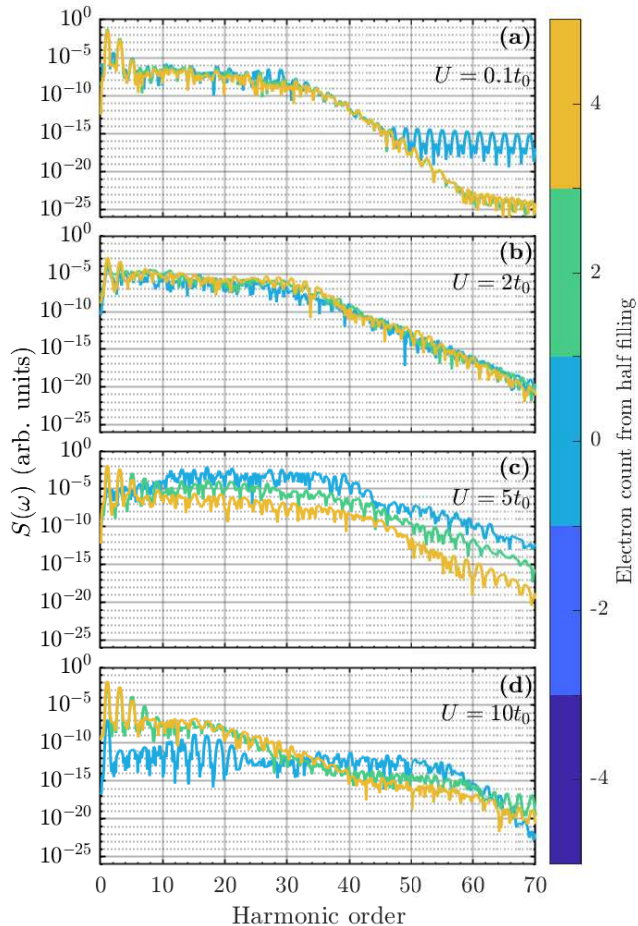}
		\caption{HHG spectra [\cref{eq:Spetrum}] for different degrees of band filling and Hubbard  $U$. (a) $U=0.1t_0$, (b) $U=2t_0$, (c) $U=5t_0$ and (d) $U=10t_0$.}
		\label{fig:Spec_ncomp}
	\end{figure}

	We begin by discussing the HHG spectra for various $U$-values and degrees of band filling, as shown in \cref{fig:Spec_ncomp}. The spectra are equal with respect to band filling around half filling, which is why only 3 of the 5 plotted degrees of band filling are visible in each panel. 
%

	First, we present some important points from \cref{fig:Spec_ncomp} and leave the discussion of their origin for later. Figures \ref{fig:Spec_ncomp} (a) and (b) show spectra for $U=0.1t_0$ and $U=2t_0$, respectively. In both cases, the changing degree of band filling has minimal impact on the spectra. However, as $U$ increases in \cref{fig:Spec_ncomp} (c), $U=5t_0$, and \cref{fig:Spec_ncomp} (d), $U=10t_0$, there are clear and very noticeable differences among the spectra for different band filling. For $U=5t_0$ a spectral enhancement of harmonic orders above the fifth is seen at half filling compared to non-half filling. This enhancement peaks at about 4-5 orders of magnitude in the plateau region between the 10th and 40th harmonic order. 
	For $U=10t_0$ half filling seems to result in a lower harmonic gain for harmonic orders lower than the 30th. However, after around the 40th harmonic, half filling seems to cause a spectral enhancement. In summary, the spectra seem to be largely unaffected by the degree of band filling, unless the correlation term is substantially larger than the hopping term, i.e., unless  $U\gg t_0$.

	\begin{figure}
		\centering
		\includegraphics[width=\linewidth]{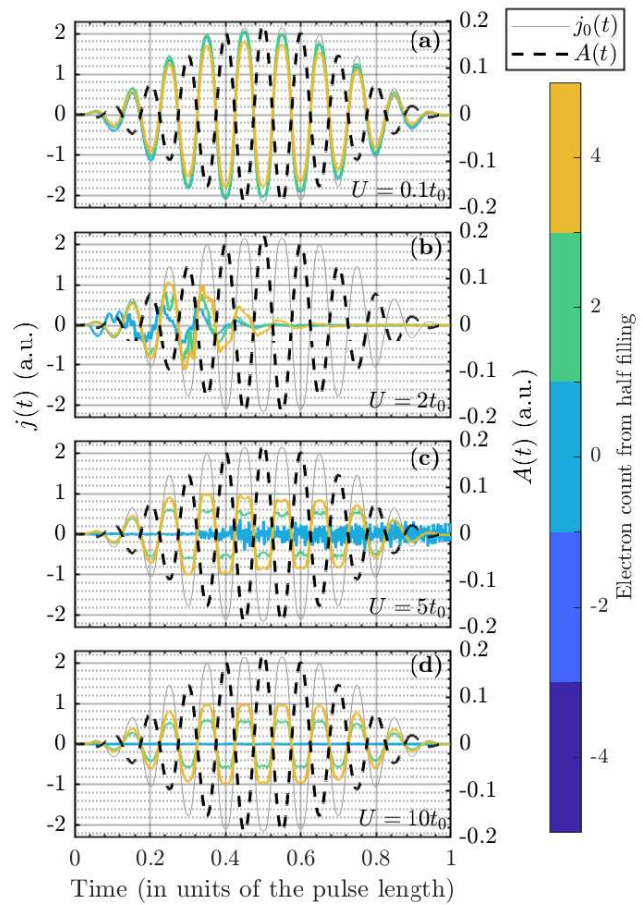}
		\caption{The current generated via \cref{eq:j}, for different degrees of band filling and Hubbard $U$. (a) $U=0.1t_0$, (b) $U=2t_0$, (c) $U=5t_0$ and (d) $U=10t_0$. In all panels the vector potential is shown by the dashed black line and the Bloch current from the half-filled lattice, $j_0(t)$ [\cref{eq:j_0}], is shown via the thin grey line.}
\label{fig:Current_ncomp}
\end{figure}
		
	\subsection{Total current results}
	To provide further insight into the spectral changes displayed in  \cref{fig:Spec_ncomp}, we show, in \cref{fig:Current_ncomp}, the currents used to generate the spectra. The increasing importance of the degree of band filling as $U$ increases, observed in \cref{fig:Spec_ncomp}, holds here too. In \cref{fig:Current_ncomp} (a), where $U=0.1t_0$, the degree of band filling has little impact. This is consistent with the analytical result which is derivable for $U=0$ where the Hubbard model reduces to a tight-binding model. This tight-binding model can be solved by employing Bloch's theorem resulting in the current given by
	\begin{align}
		j_0(t)&=aE_\text{GS}\sin\left(aA(t)\right), \label{eq:j_0}\\
		E_{GS}&=2\sum_{i=\lceil -\frac{n-1}{2}\rceil}^{\lceil \frac{n-1}{2}\rceil} \epsilon(k_i), \label{eq:EGS}\\
		\epsilon(k_i)&=-2t_0 \cos(k_i a), \label{eq:band structure}\\
		k_i&=\frac{i2\pi}{La},\quad i\in\left\{0,\pm1,\dots,\pm\frac{L}{2}-1,\frac{L}{2}\right\}.
	\end{align}
	
	Here $E_\text{GS}<0$ is the initial groundstate energy, $n$ is the number of electrons in  the lattice with a given spin, and $\epsilon(k_i)$ is the energy of the Bloch state with crystal momentum, $k_i$. We denote the rounding up operation by the $\lceil \square \rceil$ brackets. Note that for $\left| aA(t)\right| \ll 1$ the current is directly proportional to the vector potential, as is seen from a Taylor expansion of the results in \cref{eq:j_0}. This dependence on $A(t)$ is in line with the currents observed in \cref{fig:Current_ncomp} (a). 
	
	For $U=2t_0$, \cref{fig:Current_ncomp} (b), the differences between the currents for different doping levels are larger, but still show similar behavior in terms of a drop to approximately zero halfway through the simulation, which is consistent with earlier results \cite{Hansen22}.
	 
	At higher $U$, the half-filled band behaves markedly different from the others. For $U=5t_0$ [\cref{fig:Current_ncomp} (c)] the half-filled current oscillates at a much higher rate than the non-half-filled bands and rises in amplitude as the pulse amplitude increases. Furthermore, the current for the half-filled band does not decrease at the end of the pulse. For the even higher $U$ of \cref{fig:Current_ncomp} (d), $U=10t_0$, the current from the half-filled lattice has decreased to order $10^{-3}$ a.u., which is practically indistinguishable from 0 in the figure. In contrast to this, the non-half-filled lattices display currents largely similar to the $U\approx0$ currents, the most notable difference being that the peak amplitude has decreased by a factor of $\approx$2 when comparing $U=0.1t_0$ and $U=10t_0$.  
		
	\subsection{Qualitative description of the dynamics}\label{sec:explanation}
	Our understanding of the observations from \cref{fig:Spec_ncomp,fig:Current_ncomp} is best formulated in the quasiparticle picture of doublons and holons. The $U$-term of the Hubbard model punishes the creation of doublons energetically. Therefore, the groundstate will enter the configurations with minimal doublon count. For the half-filled lattice, this means the groundstate is dominated by configurations that have one electron on each site. Any electron hopping, from such configurations, will result in the creation of a doublon-holon pair. Thus virtually no dynamics can be induced from the groundstate without the creation of a doublon-holon pair. For high $U$ this results in two effects. Firstly that the current amplitude decreases significantly because transport is impeded by the cost in energy, $U$, associated with the creation of a doublon, and secondly, that the dominating mechanism is doublon-holon pair creation and annihilation as described by \cref{eq:j_ca}.
	
	If the band is not half-filled, all configurations will necessarily have one or more doublon(s) or holon(s). Therefore electrons can move about the lattice  without creating doublon-holon pairs, at no energy cost, thus enabling dynamics even in the high-$U$ limit, see \cref{fig:Current_ncomp} (d). For large $U$, the creation and annihilation of doublon-holon pairs is associated with large energy transitions and therefore with truly high-order harmonics, as will be demonstrated in further detail later. As seen in \cref{fig:Spec_ncomp} (d), the restrictions imposed by high $U$ on the dynamics in the half-filled lattice, lead to a significant drop in harmonic gain for the low to medium harmonics but to an increased gain for higher harmonics compared to the non-half-filled cases.
	
	\subsection{Non-dynamical doublon and holon measures} \label{sec:DPaf}
	In this section, we utilize the measures of \cref{eq:D,eq:Paf} to gain further understanding of the doublon-holon pair creation and annihilation throughout the simulation.
			\begin{figure}
		\centering
		\includegraphics[width=\linewidth]{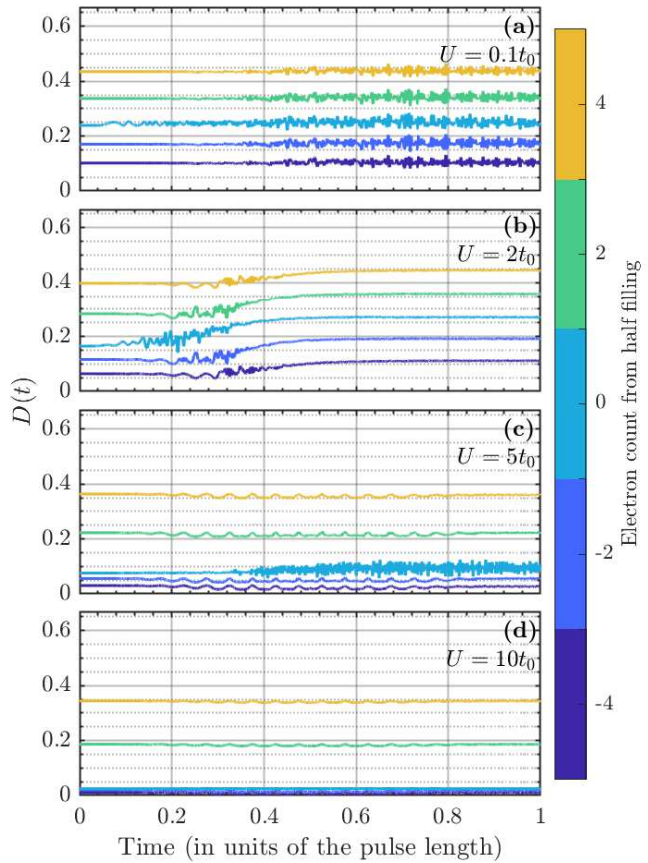}
		\caption{The $D$-measure of \cref{eq:D} for different degrees of band filling and Hubbard $U$. (a) $U=0.1t_0$, (b) $U=2t_0$, (c) $U=5t_0$ and (d) $U=10t_0$.}
		\label{fig:D}
	\end{figure}
	\subsubsection{$D$-measure}\label{sec:D}
	Figure~\ref{fig:D} shows $D(t)$, see \cref{eq:D}, for different degrees of band filling and $U$. We note immediately from all panels that removing or adding electrons to the lattice will change the number of doublons in the lattice. To estimate the impact of varying band filling on $D(t)$ we calculated the $D$-measure from the state in which every possible electron configuration is equally likely. This resulted in the following values, with 4 electrons removed from half filling: $D\simeq 0.111$, with 2 electrons removed $D\simeq 0.174$, at half filling $D\simeq 0.25$, with 2 electrons added $D\simeq 0.340$, and with 4 electrons added $D\simeq0.444$. In \cref{fig:D} (a) the $D$-measure is plotted for $U=0.1t_0$. The $D$-values are close to constant and very close to the values given above for equal probability of all electron configurations. This makes sense as for $U\ll t_0$ the system's eigenvalues are close to the Bloch states, which are the eigenstates for $U=0$. The Bloch states are completely localized in $k$-space, so therefore completely delocalized in real space. This delocalization causes each electron configuration to be approximately equally probable. We note from Figs. \ref{fig:D} (b), (c), and (d) that the initial groundstate, has decreasing $D$-values with increasing $U$ and below the values from states with equal probability of all electron configurations. This is a result of the increasing energetic cost to create doublons. Minimizing the energy of the  state implies minimizing the number of doublons, and therefore the probability of observing configurations with a high number of doublons. In \cref{fig:D} (b), with $U=2t_0$, there is a noticeable increase in the $D$-values over the time interval $0.2-0.5$ pulse lengths. This increase is not present for lower or higher $U$-values. This can be explained by considering the correlation between the number of doublons in and the energy of a given state, as well as general features of the energy spectrum. For low $U$ the energy of a state is largely determined by the hopping term of the Hamiltonian [\cref{eq:Hamiltonian}], whereas for higher $U$ it is determined by the $U$-term. As mentioned, Bloch states, the eigenstates of the hopping-term, are delocalized in real-space and as a result, there is little to no correlation between the energy and the number of doublons in the state. On the other hand, the expectation value of the $U$-term depends only on the $U$-value and the number of doublons in the state, resulting in complete correlation between the number of doublons in the state and the energy from the $U$-term. This means for low $U$, exciting the system, i.e., increasing the energy of the system, does not increase the number of doublons in the state, causing $D$ to remain virtually constant. This explains why $D$ is approximately constant in \cref{fig:D} (a). For higher $U$, see Figs.\ref{fig:D} (c) and (d) with $U=5t_0$ and $U=10t_0$, respectively, the $D$-values are also approximately constant, despite the high degree of correlation between the energy and doublon count. This can be understood by considering the energy gaps between the eigenvalues of the system. For $U=0$ many of the states are highly degenerate; see Ref.~\cite{Hansen22} for a discussion. That degeneracy is removed when $U$ becomes non-zero. As $U$ increases further the energy gaps between the states increase steadily, resulting in relatively large energy gaps. Since the pulse used is the same throughout this work, the degree of system excitation decreases as $U$ increases. So due to the minimal degree of excitation, the $D$-values change minimally in both \cref{fig:D} (c) and (d), despite the correlation between energy and $D$-value. 
	
	For $U\rightarrow\infty$, the $D$-measure attains the value corresponding to the configurations with the minimal number of doublons. If the lattice is half filled or less, then there is at minimum one site per electron and $D=0$, i.e., states with no doublons, can be achieved. When the lattice is more than or half filled each added electron adds another doublon, causing an increase in $D$ of $\frac{1}{L}\simeq0.083$. This is in line with the $D$-values of \cref{fig:D} (c) and (d). Finally, we note that the oscillations observed in the $D$-values of Figs. \ref{fig:D} (c) and (d) reach a local maximum when the electric field from the pulse (not shown) peaks, i.e., when the rate of excitation is maximal. 
			\begin{figure}
		\centering
		\includegraphics[width=\linewidth]{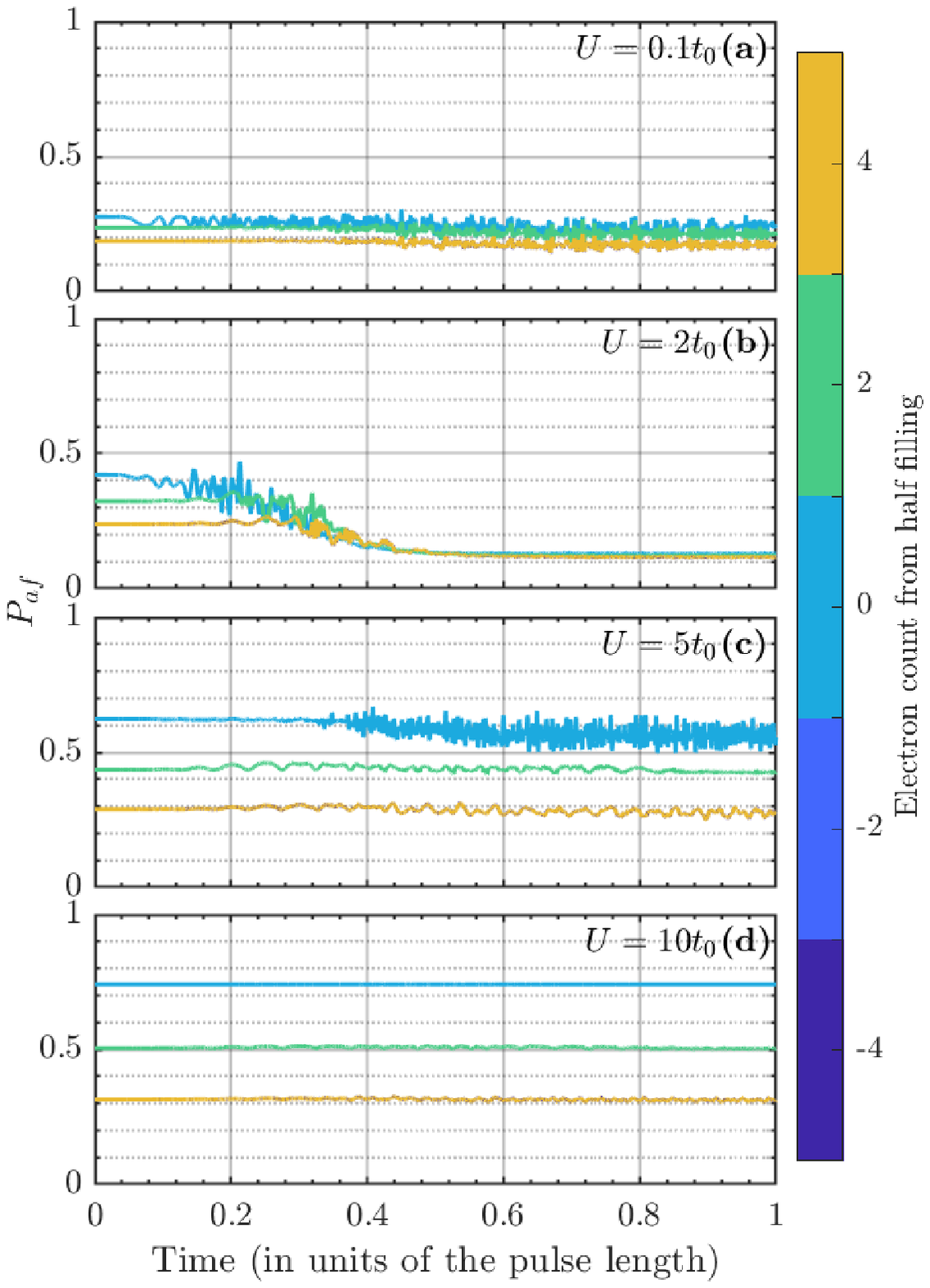}
		\caption{The $P_\text{af}$ measure of \cref{eq:Paf}, for different degrees of band filling and Hubbard $U$. (a) $U=0.1t_0$, (b) $U=2t_0$, (c) $U=5t_0$ and (d) $U=10t_0$.}
		\label{fig:Paf}
	\end{figure}

	\subsubsection{$P_\text{af}$-measure}\label{sec:Paf}
 	Here we discuss the $P_\text{af}$-measure of \cref{eq:Paf}, which is shown in \cref{fig:Paf}. As $P_\text{af}$ is heavily dependent on the number of electrons in the lattice it can be difficult to compare $P_\text{af}$-values for different degrees of band-filling. In order to give an idea of those changes in the $P_\text{af}$-values, we give the $P_\text{af}$ values for states with equal probability of observing each electron configuration. At half filling $P_\text{af}\simeq 0.149$, with 2 electrons added or removed from half filling $P_\text{af}\simeq 0.141$, and with 4 electrons added or removed from half filling $P_\text{af}\simeq 0.118$. 
Firstly, we note that the $P_\text{af}$-values are virtually constant at all times for $U=0.1t_0$, $U=5t_0$, and $U=10t_0$, \cref{fig:Paf} (a), (c), and (d) respectively.  This is for the same reasons as those explaining the equivalent observation on the $D$-values, presented in Sec. \ref{sec:D}, i.e., due to a low correlation between the energy, doublon count, and antiferromagnetic ordering in the initial groundstate for low $U$, and large energy gaps between the eigenstates, leading to a low degree of excitation, for high $U$. We note, however, that the $P_\text{af}$-values decrease in \cref{fig:Paf} (b), simultaneously with the increase in $D$-values, between $0.2$ and $0.5$ pulse lengths. We understand the decrease in terms of the increasing number of doublons resulting in fewer electrons and sites having the possibility of being antiferromagnetically ordered, thus decreasing $P_\text{af}$. 
	\subsubsection{Summary regarding the $D$- and $P_\text{af}$-measures} \label{sec:D_Paf_summary}
	We conclude from the observations of the previous two subsections that for low $U$, doublon-holon pairs are being created and annihilated, but as there is little energy involved with this process, the rates of creation and annihilation are very close to equal. We note also, based on the low $P_\text{af}$ values, see \cref{fig:Paf} (a), that the creation and annihilation of doublon-holon pairs correspond to a smaller portion of the dynamics than for higher $U$. For higher $U$ the system is less dynamical overall, as indicated by the currents shown in \cref{fig:Current_ncomp}. The higher $P_\text{af}$ and increasing $D$-measure in this limit indicates that doublon-holon-pair-creating or -annihilating transitions are more prevalent. As $U$ becomes significantly larger than $t_0$ the interpretation of the results begins to depend significantly on the degree of band filling. In the half-filled case there are virtually no dynamics left in the system, but the transitions that do happen are virtually exclusively of the doublon-holon pair creating or annihilating type. Whereas for the non-half-filled lattices the existence of either doublons or holons in the groundstate implies that transitions at no $U$-induced energy cost are possible. The final configurations of such transitions may act as intermediary configurations to the creation of doublon-holon pairs and thus result in more doublon-holon pair creating and annihilating transitions. As we shall now see all of the conclusions and indications in this section are consistent with behavior of the currents $j_\text{ca}(t)$ and $j_\text{hop}(t)$ and the associated spectra.

	\begin{figure}
		\centering
		\includegraphics[width=\linewidth]{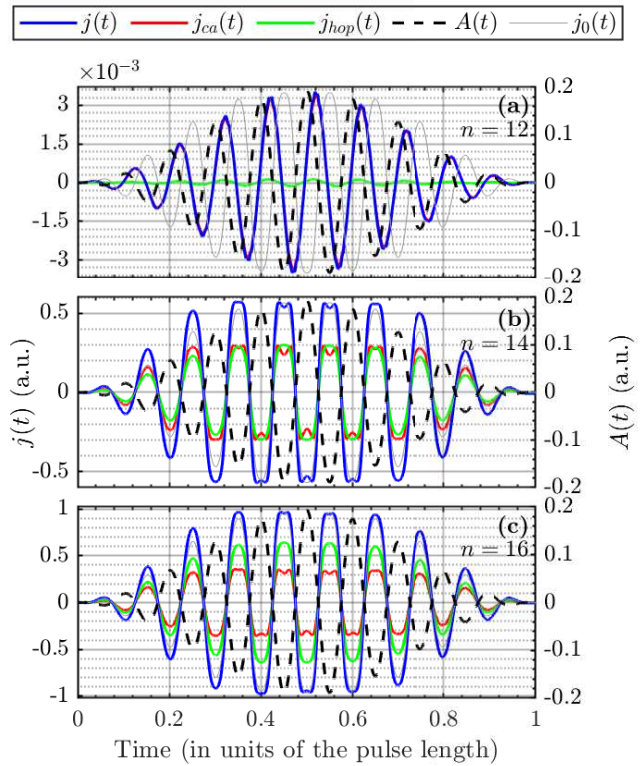}
		\caption{The creation and annihilation current, $j_\text{ca}(t)$, from \cref{eq:j_ca}, and the hopping current, $j_\text{hop}(t)$, from \cref{eq:j_hop}, for the $U=10t_0$ system for lattices with $n$ electrons. Half filling corresponds to $n=L=12$. The total current is shown by the full, blue curve, and the vector potential is shown by the black, dashed line. Finally the scaled down Bloch-current is shown by the thin, grey line. The scaling factors are given by ${\max(\left|j(t)\right|)}/{\left|2*a*E_\text{GS}\right|}$, (a) $\simeq1.9*10^{-3}$, (b) $\simeq 0.31$, (c) $\simeq 0.52$.}
		\label{fig:j_ca_hop}
	\end{figure}

	\subsection{Quasi-particle currents} \label{sec:j_ca_hop}
	To support the explanation of \cref{sec:explanation}, regarding doping effects on the dynamics, and verify the points from \cref{sec:D_Paf_summary}, we show $j_\text{ca}(t)$ [\cref{eq:j_ca}] and $j_\text{hop}(t)$ [\cref{eq:j_hop}] for $U=10t_0$ in \cref{fig:j_ca_hop}. We only plot the currents for lattices which are at least half filled, due to the earlier mentioned symmetry.
In \cref{fig:j_ca_hop} (a) the currents in the half-filled lattice are shown. We note that the majority of the current arises from the creation or annihilation of doublon-holon pairs, $j_\text{ca}(t)$. This finding is consistent with the explanation described above, as for $U=10t_0$ the groundstate will be dominated by the antiferromagnetically ordered, doublon-holon-pair-free configurations. From such configurations any electron hopping will lead to the creation of a doublon-holon pair. That is unlikely due to the associated energy cost, $U$. Note also that, in this case, the current is only of order $10^{-3}$ a.u. For a lattice with 14 electrons, \cref{fig:j_ca_hop} (b), the current arising from the hopping of the doublons and holons approximately equals the current arising from creation or annihilation of said quasiparticles. It is worth noting that with 14 electrons in the lattice the groundstate is dominated by configurations with 2 doublons. This means there are significantly more ways for an electron to hop resulting in the creation of a doublon-holon pair than in a doublon hopping. Figure \ref{fig:Paf} verifies this by showing that the probability of observing antiferromagnetic ordering is consistently higher for states with 2 electrons added relative to half filling compared with the probability from half-filled states. Note that the result of annihilating a doublon-holon pair is antiferromagnetic ordering, and inversely that antiferromagnetic ordering is needed to facilitate doublon-holon-pair-creating transitions, see Fig.~\ref{fig:illustration}. The currents in \cref{fig:j_ca_hop} are all for $U=10t_0$ corresponding to \cref{fig:Paf} (d). We note from \cref{fig:Paf} that the probability of observing antiferromagnetic ordering generally increases with $U$. That trend is reasonable since antiferromagnetically ordered configurations do not contain doublons and  the impact on $P_\text{af}$ of removing or adding electrons to the lattice increases similarly, as seen from the sizes of the gaps between the lines in \cref{fig:Paf}. If there are more possible transitions from the groundstate that may result in the creation of a doublon-holon pair, than transitions which move doublons or holons, then everything else being equal, $j_\text{ca}(t)>j_\text{hop}(t)$. So that $j_\text{ca}(t)\approx j_\text{hop}(t)$, for this lattice, indicates that hopping transitions are more likely than creation or annihilation transitions relative to the number of lattice gaps that enable the given type of transition.
	
	 With 16 electrons in the lattice, \cref{fig:j_ca_hop} (c), the hopping current is larger than the creation or annihilation current at all times, consistent with the larger number of doublons in the groundstate. 
	 
	 We note that the currents in \cref{fig:j_ca_hop} (a) are out of phase with the Bloch current, $j_0(t)$ [\cref{eq:j_0}]. The currents peak when the electric-field peaks, instead of when the vector potential peaks. That is because the groundstate is dominated by the anti-ferromagnetically ordered electron configurations. Any transition from such a configuration results in the creation of a doublon-holon pair, which, for high $U$, is most likely at high laser intensity, and therefore peak electric-field. Inversely the currents in the non-half-filled lattices are in phase with the Bloch current. That is due to the free electrons associated with the ever-present doublons or holons which, qualitatively, are Bloch like, i.e., free to move about the lattice, and only interacting with other particles via Pauli's exclusion principle.
	 
	 Finally, we note that the current amplitude scale approximately linearly with the number of electrons added or removed from the lattice relative to half filling, i.e., with the number of free carriers in the form of doublons or holons. This scaling is consistent with the Drude model.

\begin{figure}
		\centering
		\includegraphics[width=\linewidth]{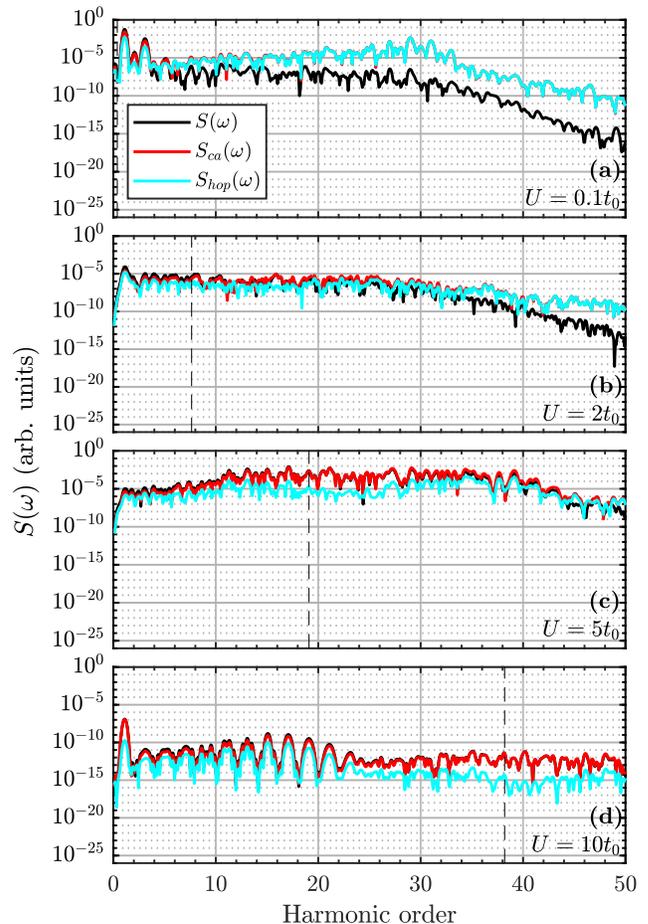}
		\caption{The  creation and annihilation, $S_\text{ca}(\omega)$, and hopping, $S_\text{hop}(\omega)$ spectra [\cref{eq:S_ca and S_hop}] from the half-filled lattice with various values of $U$, (a) $U=0.1t_0$, (b) $U=2t_0$, (c) $U=5t_0$ and (d) $U=10t_0$. The dashed, black vertical lines indicate the $U$-value in terms of harmonic orders.  }
		\label{fig:S_ca_hop}
	\end{figure}

	\subsection{Quasi-particle spectra}
	In order to obtain more insight into the link between the two types of dynamics associated with, $\hat{j}_\text{ca}(t)$ [\cref{eq:j_ca}] and $\hat{j}_\text{hop}(t)$ [\cref{eq:j_hop}], we show spectra generated from each current type alone in \cref{fig:S_ca_hop}. All the plots are from the half-filled lattice, as the crucial points are easiest to see from this case. We note that the sum of the spectra from these two currents neither does nor should add up to the spectrum generated from the total current due to interference terms from the norm-square in \cref{eq:Spetrum}; see text after Eq.~\eqref{eq:S_ca and S_hop}. It is, however, still reasonable to expect that general trends concerning which harmonics are generated from which mechanism can be gleaned from this analysis. We note from \cref{fig:S_ca_hop} (a), $U=0.1t_0$, that there are little to no differences between the hopping and creation and annihilation spectra for that relatively small $U$-value, to the point where $S_\text{ca}(\omega)$ and $S_\text{hop}(\omega)$ overlap after the 11th harmonic order. The only differences worth noting are the slightly larger 1st, 3rd, and 5th order harmonic peaks from $S_\text{ca}(\omega)$ compared to $S_\text{hop}(\omega)$. For $U=2t_0$, \cref{fig:S_ca_hop} (b), the differences are more noticeable, however, still rather small. They peak at around the 10th to 20th harmonic orders. For $U=5t_0$, $S_\text{ca}(\omega)$ is multiple orders of magnitude higher at harmonic orders 15-30. Finally for $U=10t_0$, $S_\text{ca}(\omega)$ is larger than $S_\text{hop}(\omega)$ across the entire spectrum, but the difference is largest for harmonic orders 35-45. We conclude two things from these observations. Firstly, the increasing difference between the two spectra for increasing $U$ is likely a result of the groundstate being more and more dominated by the doublon-free configurations, from which the only transitions available result in the creation or annihilation of a doublon-holon pair, as \cref{fig:S_ca_hop} is for the half-filled lattice. This overall enhancement of $S_\text{ca}(\omega)$ is therefore expected to decrease as the degree of band filling moves away from half filling. This is also observed from the non-half-filled spectra (not shown). The second observation from \cref{fig:S_ca_hop} is that the difference between $S_\text{ca}(\omega)$ and $S_\text{hop}(\omega)$ peaks at around the harmonic order corresponding to the $U$-value. This observation is a result of transitions that result in the creation or annihilation of a doublon-holon pair absorbing or releasing energy of order $\approx U$ plus the energy of said transition in the correlation-free case. This insight also explains why this difference only becomes distinct for $U>4t_0$, as $4t_0\simeq 15\omega_L$ is the bandwidth from the correlation-free case, see \cref{eq:band structure}. When $U<4t_0$ transitions, resulting in movement of doublons or holons can correspond to energies of $U$, which is not the case for $U>4t_0$. Therefore when $U>4t_0$ the main way for electrons to transition under emission of a photon with energy $\approx U$ is through annihilation of a doublon-holon pair resulting in a noticeable difference between $S_\text{ca}(\omega)$ and $S_\text{hop}(\omega)$. We note that transitions resulting in the creation or annihilation of a doublon-holon pair should be able to generate energies between $\max(U-4t_0,0)$ and $U+4t_0$, which explains why the region of relative enhancement in Figs.~\ref{fig:S_ca_hop} (c) and (d) is $\approx8t_0\simeq 30$ harmonic orders wide.
	
	When going beyond the half-filled case, the $P_\text{af}$-value decreases, as seen in \cref{fig:Paf}. This means the creation and annihilation dynamics become more and more insignificant resulting in the tendencies discussed above, regarding enhancement of $S_\text{ca}(\omega)$ compared to $S_\text{hop}(\omega)$, becoming less and less significant. For 16 electrons in the lattice there are minimal differences between $S_\text{ca}(\omega)$ and $S_\text{hop}(\omega)$ across the spectrum in general and the observed extra enhancement around harmonic orders corresponding to the $U$-value disappears.

	\section{summary and conclusion}\label{sec:conclusion}
	In this work, we have used the Hubbard model to study HHG beyond the assumption of half filling, corresponding to a situation where the correlated material is doped. We have done this to begin answering how HHG operates from highly doped and correlated materials. In previous studies of HHG using the Hubbard model for the half-filled case, it was shown that increasing the Hubbard $U$ leads to increased gain in the HHG spectra for $U\leq 5t_0$ \cite{Silva2018}. On top of that a parallel to the 3-step model was developed in the half-filled case based on the creation and annihilation of doublons and holons \cite{Murakami2021}. 
	With these facts in mind, we set out to answer the following questions:  (i) What effects does doping have on the HHG spectra? (ii) How do the dynamics of non-half-filled bands differ from half-filled bands? (iii) How do the changes in the dynamics and HHG spectra relate to one another?
	It was found that going beyond half filling has little to no effect on the spectra for low $U$, but that for higher $U$ there are marked differences. For $U=5t_0$ the half-filled lattice showed significantly higher high-harmonic gain than the non-half-filled lattices. For higher $U$ the half-filled lattice droped to be the lattice with least harmonic gain for low to medium harmonic orders but still gave the highest gain for high-harmonic orders. This finding may be rationalized in terms of the high energy needed to produce doublon-holon pairs. We explained this by splitting the dynamics into two mechanisms: (i) movement of the doublons or holons, the parallel to intraband dynamics, and (ii) creation and annihilation of doublon-holon pairs, the parallel to interband dynamics. For high $U$-values the groundstate of the Hubbard model is dominated by the configurations with the highest probability of observing antiferromagnetic ordering, see \cref{fig:Paf}.  For half-filling this results in every hop of the electrons creating a doublon-holon pair. For high $U$, this pair creation requires a lot of energy, resulting in minimal dynamics and thus minimal harmonic gain. For lower $U$, the lower energy cost of creating a doublon-holon pair means that relatively more dynamics associated with the creation and annihilation of doublon-holon pairs take place. When the lattice is not half filled, all configurations, and by extension, the groundstate will contain either doublons or holons. As a result, regardless of the size of the $U$-term, dynamics associated with the movement of doublons or holons can take place at no extra $U$-induced energy cost. This absence of an energy penalty results in the spectra being significantly less affected by the $U$-term and as a result, higher high-harmonic gain in the high-$U$ limit compared with the half-filled case was observed. We substantiated this explanation by investigating the currents associated with the two mechanisms in the high-$U$ limit. In the half-filled case, there is significantly less current than in the non-half-filled cases, and it is dominated by the creation and annihilation current. In the non-half-filled cases, the current associated with the movements of the doublons or holons becomes increasingly important, with the total current rising in amplitude accordingly. We analyzed spectra generated from the hopping and creation and annihilation currents only. We found that in the half-filling case, increasing $U$ leads to the majority of the spectral gain being formed through creation and annihilation dynamics. It was also found that regardless of the degree of band filling the creation and annihilation spectrum had higher gain around harmonic orders corresponding to the $U$-value considered, particularly when the $U$-value is larger than the band width of the correlation-free Hamiltonian, i.e., $4t_0$. This relative enhancement was over a relatively large number of harmonic orders, a result of the creation and annihilation transitions resulting in absorptions and emissions between $\sim |U-4t_0|$ and $\sim |U+4t_0|$. Finally, we showed the expectation value of the $D$-measure, which similarly to the current analysis outlined above, was in line with the explanation presented. 
	
	In conclusion, the present study shows that the degree of band filling is of immense importance to the gain in HHG for highly correlated materials. In cases with doping away from the half-filling situation, the doping-induced enhancement of the HHG spectra extends to the lowest harmonic orders and can be enhanced by as much as 10 orders of magnitude. The significant enhancement is a very clear signature of the sensitive, ultrafast response of correlated electron dynamics  to extreme nonperturbative driving by intense laser pulses.
		
	\acknowledgments
	This work was supported by the Independent Research Fund Denmark (Grant No. 1026-00040B).

\end{document}